\documentclass{moriond}

\def\be{\begin{equation}}
\def\ee{\end{equation}}
\def\bea{\begin{eqnarray}}
\def\eea{\end{eqnarray}}

\newcommand{\Photo}{}

\begin{document}
\hfill CERN--PH--TH/2015--146

\vspace*{4cm}
\title{Introduction to Heavy Flavours}

\author{ F. MAHMOUDI }

\address{Universit{\'e} de Lyon, Universit{\' e} Lyon 1, F-69622 Villeurbanne Cedex, France;\\
Centre de Recherche Astrophysique de Lyon, CNRS, UMR 5574, Saint-Genis Laval Cedex, F-69561, France;
Ecole Normale Sup{\'e}rieure de Lyon, France\\
and\\
CERN Theory Division, Physics Department, CH-1211 Geneva 23, Switzerland}

\maketitle\abstracts{
We present a short review of heavy flavours and the main challenges given the recent experimental developments.}

\section{Introduction}

Many fundamental questions in particle physics are related to flavour. Among the most important open issues one can mention the hierarchy of quark masses, the absence of flavour changing neutral currents, the pattern of mixing angles of quarks, the origin of the baryon asymmetry in the Universe and the number of flavours and quarks. While there have been several attempts to provide answers to these questions, by considering for example continuous or discrete flavour symmetries, extra-dimension models or compositeness, no definite picture is yet attained. The quark sector is nevertheless well described through the Cabibbo-Kobayashi-Maskawa (CKM) formalism.

In this review, we will first briefly address the CKM mechanism and in particular the determination of $V_{ub}$ for which there are discrepancies in the measurements. We will then move to indirect search for New Physics (NP) with the two main actors, which are tests of CP violation and studies of rare decays.

\section{CKM formalism}

\begin{figure}[h!]
\begin{center}
\includegraphics[width=0.8\linewidth]{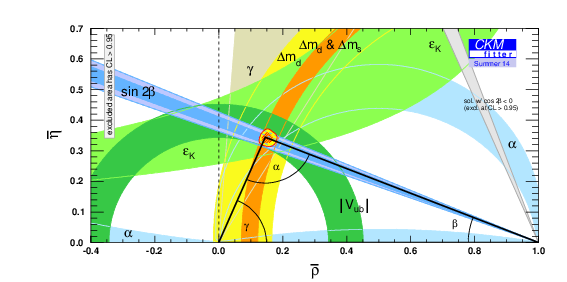}
\end{center}
\caption[]{Unitarity triangle in the $\bar\rho=\rho(1-\lambda^2/2)$ vs. $\bar\eta=\eta(1-\lambda^2/2)$ parameter plane \cite{CKMfitter2014}.}
\label{fig:ckm}
\end{figure}

In the Standard Model (SM), the mixing within the three generations of quarks is described by the CKM matrix. In the Wolfenstein parametrisation, it can be written as 
\begin{equation}
V_{\rm CKM} = \left( \begin{array}{ccc}
      V_{ud} & V_{us} & V_{ub}\\
      V_{cd} & V_{cs} & V_{cb}\\
      V_{td} & V_{ts} & V_{tb}\end{array} \right)
      = \left( \begin{array}{ccc}
      1-\lambda^2/2 & \lambda & A \lambda^3(\rho + i \eta)\\
      -\lambda & 1-\lambda^2/2 & A \lambda^2\\
      A \lambda^3(1 - \rho - i \eta) & - A \lambda^2 & 1\end{array} \right) \label{eq:ckm}
\end{equation}
which contains three CP conserving parameters and one CP violating phase, and is unitary. The results of two decades of Babar, Belle and LHCb in measuring the CKM parameters is summarised in Fig.~\ref{fig:ckm}, which confirms the unitarity of the CKM matrix, and shows that the CKM paradigm is fully consistent with the data.

\begin{figure}[h!]
\begin{center}
\includegraphics[width=0.35\linewidth]{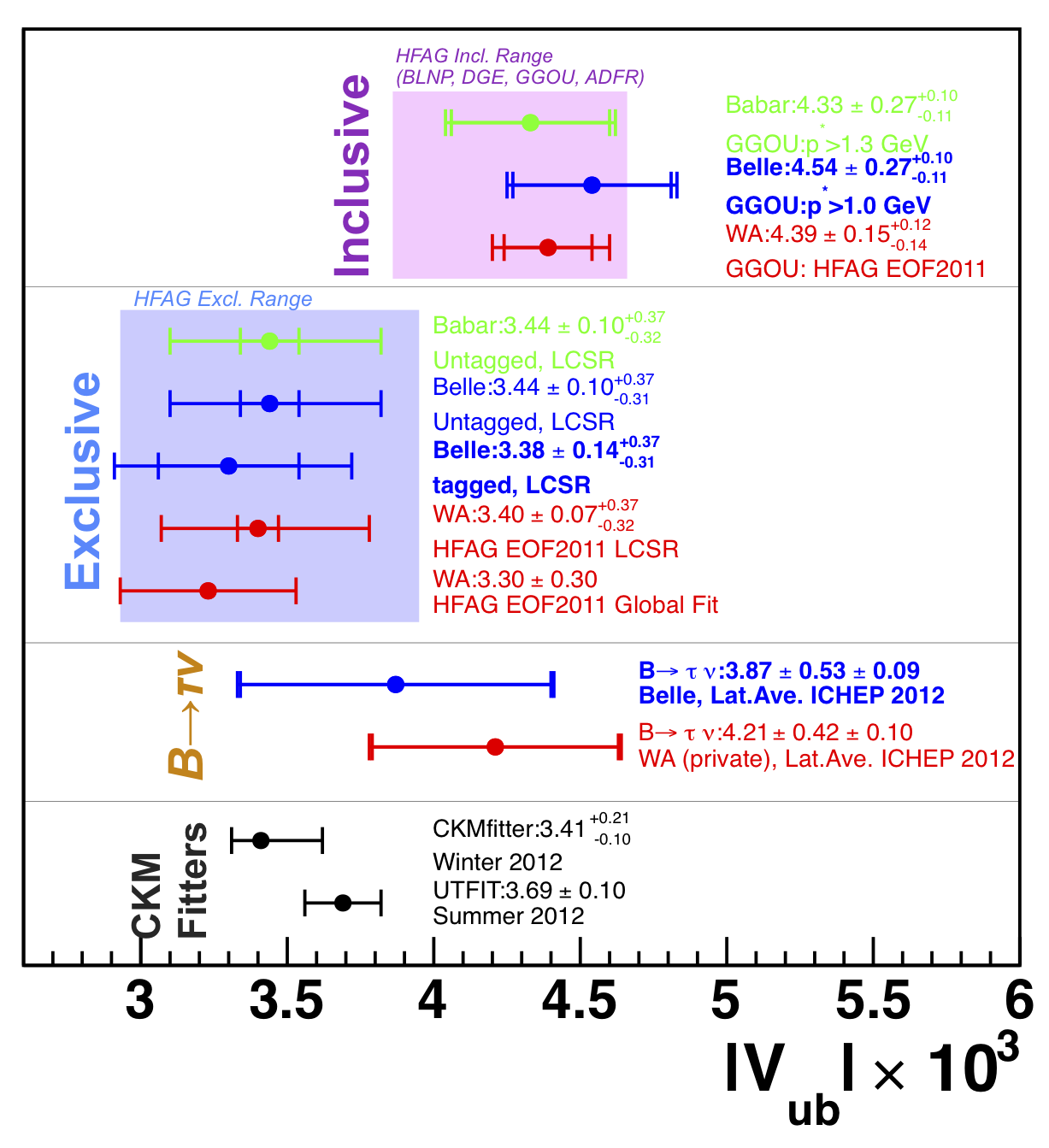}\hspace*{1.5cm}\includegraphics[width=0.45\linewidth]{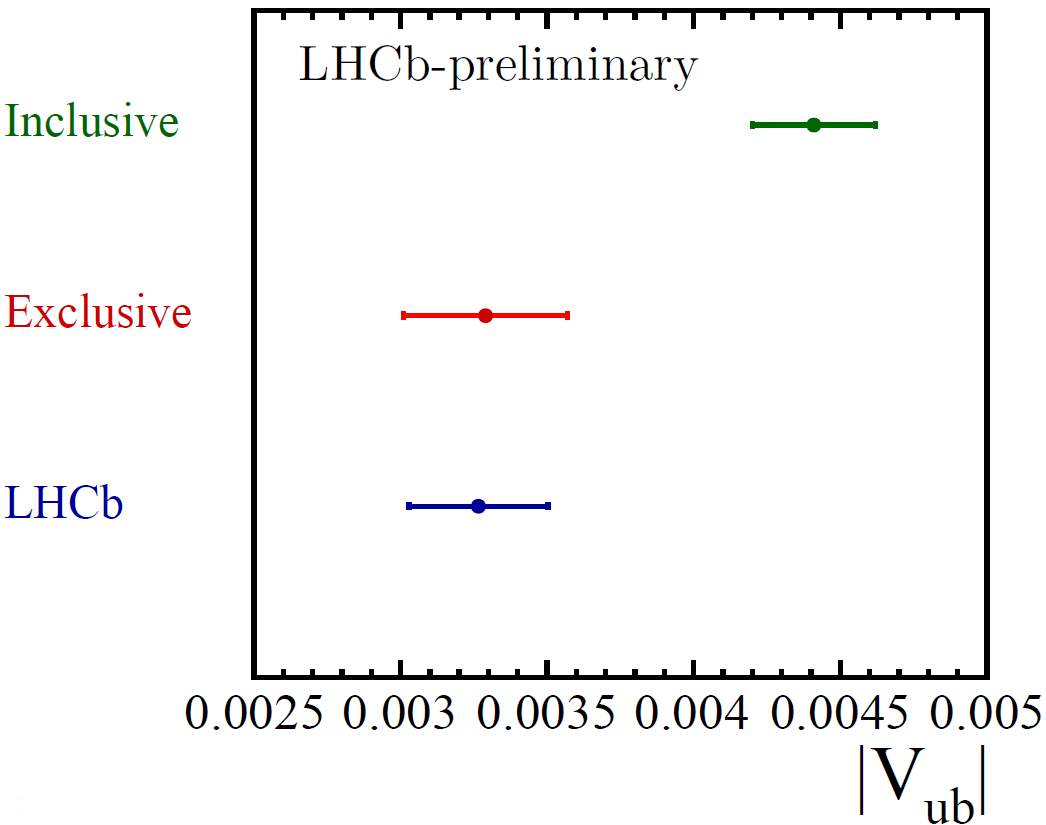}
\end{center}
\caption[]{Summary of the different measurements of $|V_{ub}|$ at $B$ factories~\cite{Urquijo:2013npa} (left panel), and comparison with the LHCb result~\cite{Aaij:2015vza} (right panel).}
\label{fig:vub}
\end{figure}

One of the least known parameters of the CKM matrix is $V_{ub}$. 
Measurement of $|V_{ub}|$ is very challenging, but the precision is reduced to $\sim 10\%$ at $B$ factories.
The two main ways to measure $|V_{ub}|$ are based on inclusive semi-leptonic decays ($B\to X_u \ell \nu$) and exclusive semi-leptonic decays ($B\to \pi\ell \nu$).
Both approaches provide independent measurements of $|V_{ub}|$. There is currently a discrepancy in the central values of about 3$\sigma$, but they have roughly the same precision. Both methods can be employed at a high luminosity $B$ factory, and the experimental error in the determination of $|V_{ub}|$ will decrease with increasing integrated luminosity and improved analysis techniques.
The inclusive and exclusive determinations of $|V_{ub}|$ have independent theoretical errors~\cite{Akeroyd:2010qy}.
Concurrent reduction of this theoretical error is however more challenging. A summary of the current measurements is provided in Fig.~\ref{fig:vub}.

The current combination of $|V_{ub}|$ from inclusive decays using nine independent measurements of $B\to X_u\ell \nu$ gives~\cite{Agashe:2014kda}:\\
\begin{equation}
|V_{ub}|=\bigl(4.41\pm 0.15({\rm exp}) ^{+0.15}_{-0.17}({\rm theory})\bigr)\times 10^{-3}\;,
\end{equation}
where the experimental error is of the same order as the theoretical error.
Conversely, the world average measurement of $|V_{ub}|$ from exclusive decays is~\cite{Agashe:2014kda}:
\begin{equation}
|V_{ub}|=(3.28\pm 0.29)\times 10^{-3}\;.
\end{equation}

Very recently, LHCb has also measured $|V_{ub}|$ using baryonic decays 
\cite{Aaij:2015bfa}. This is an important breakthrough, as it was thought for long that the measurement of $|V_{ub}|$ is impossible at hadron colliders.
The most promising channels at LHCb are $\Lambda_b \to p \,\mu \nu$ and $B_s \to K^+ \mu \nu$.
$\Lambda_b \to p \,\mu \nu$ is more favourable due to smaller background with protons.
Both decays have branching ratios of about $10^{-5}-10^{-4}$, and precise lattice calculations of $\Lambda_b \to p, \Lambda_{(c)}$ form factors are now available~\cite{Detmold:2015aaa}. Experimentally, such decays are challenging because of neutrino reconstruction. The main background comes from decays involving $V_{cb}$. The determination of $|V_{ub}|$ from $\Lambda_b \to p \,\mu \nu$ gives:
\begin{equation}
|V_{ub}|=\bigl(3.27\pm 0.15({\rm exp}) \pm 0.17({\rm theory}) \pm 0.06(|V_{cb}|)\bigr)\times 10^{-3}\,,
\end{equation}
which is 3.5$\sigma$ below the inclusive measurement but agrees well with current exclusive world average, as can be seen in the right panel of Fig.~\ref{fig:vub}.

$V_{ub}$ is a particularly important CKM parameter since it is also related to the question of CP violation.

\section{CP violation}

CP violation is a key concept to explain the baryon asymmetry in the Universe. Since the CKM matrix is the only source of CP violation in the SM, any sign of extra CP violation would point to New Physics.
As can be seen from Eq.~(\ref{eq:ckm}), $V_{ub}$ and $V_{td}$ are the only CP violating CKM parameters in the SM.
$V_{ub}$ is probed directly in $b\to u$ transitions. $V_{td}$ on the other hand involves a top quark and therefore is probed indirectly, for example in $B_d$ mixings.

\subsection{CP violation in $B_{(s)}$ mixings}

The study of the $B_{(s)}-\bar B_{(s)}$ oscillations allows for indirect probe of CP violation in $V_{td}$ ($V_{ts}$).
$B$ and $\bar B$ have different properties which come from the difference in the heavy and light eigenstates.
Two important observables are the mass difference and the decay rate difference between the two eigenstates,
which can be written as
\begin{equation}
\Delta M \equiv M_H - M_L = 2 |M_{12}|\;, \hspace*{1.cm} \Delta \Gamma \equiv \Gamma_L - \Gamma_H = 2 |\Gamma_{12}| \cos \phi_q\; , 
\end{equation}
where $M_{12}$ and $\Gamma_{12}$ are respectively sensitive to the heavy (i.e. top quarks and New Physics) and light internal particles.
The deviation from the SM is parametrised by $\Delta_q \, (q = d, s)$ in a generic way: 
\begin{equation}
 M_{12,q} = M_{12,q}^{SM} \times \Delta_q \;.
\end{equation}
A complex $\Delta_q$ implies new source of CP violation, and $\phi_q \equiv \arg(\Delta_q)$ is the CP violating weak phase.

$B_{(s)}$ mixings are particularly studied using the semi-leptonic asymmetries (for example $B_{(s)} \to D_{(s)} \ell \nu$), which allows us to measure the quantity $a_{\rm sl}$, related to the imaginary part of the ratio of $\Gamma_{12}$ over $M_{12}$, or $\phi_q$ and $\Delta_q$:
\begin{equation}
a_{\rm sl}^q = {\rm Im}\left(\frac{\Gamma_{12,q}}{M_{12,q}}\right) = \left(\frac{|\Gamma_{12,q}|}{  |M_{12,q}|}\right) \, \frac{\sin(\phi_q^{SM} + \phi_q)}{|\Delta_q|}\;.
\end{equation}
The measurements of $a_{\rm sl}^d$ and $a_{\rm sl}^s$ by LHCb~\cite{Aaij:2013gta,Aaij:2014nxa} and $a_{\rm sl}^d$ by the $B$ factories~\cite{Amhis:2014hma} are well in agreement with the SM predictions~\cite{Lenz:2011ti}, while the measurement of these two quantities in combination with the dimuon charge asymmetry by the D0 experiment shows a deviation of more than 3$\sigma$ from the SM prediction~\cite{Abazov:2012zz}.

\subsection{CP violation in $B_{(s)}$ decays}

One way to probe more directly CP violation is to consider decays into neutral mesons. For example by comparing $B^- \to D^0 K^-$ and $B^- \to \bar D^0 K^-$ one can directly probe the difference between the $D^0$ and the $\bar D^0$. These decays correspond to different diagrams which are either sensitive to the CP conserving CKM parameters $V_{cb}$ and $V_{us}$, or to $V_{ub}$ and $V_{cs}$. Other channels can also be studied such as $B \to \phi K^{(*)}, K \pi $, $B_s \to KK, \pi\pi, K\pi, \phi\phi, J/\psi\, \phi$.

The way to derive a CP asymmetry is different when the initial meson is charged or neutral, because of the oscillations. For the charged mesons, the measurement consists of studying directly the decay with a positive charge and the decay with a negative charge, and computing a CP asymmetry, leading to a direct determination of this asymmetry:
\begin{equation}
A_{CP} \equiv \frac{\Gamma(M^+ \to f^+) - \Gamma(\bar M^- \to f^-)}{\Gamma(M^+ \to f^+) + \Gamma(\bar M^- \to f^-)}\;.
\end{equation}
For neutral mesons, because of the oscillations, the determination is more complicated and is mostly indirect. The CP asymmetry is defined similarly to the case of charged mesons but it is time dependent because of the oscillations:
\begin{equation}
A_{CP}(t) \equiv \frac{\Gamma(M^0 \to f ; t) - \Gamma(\bar M^0 \to f ; t)}{\Gamma(M^0 \to f ; t) + \Gamma(\bar M^0 \to f ; t)}\;.
\end{equation}

\begin{figure}[t!]
\begin{center}
\includegraphics[width=0.45\linewidth]{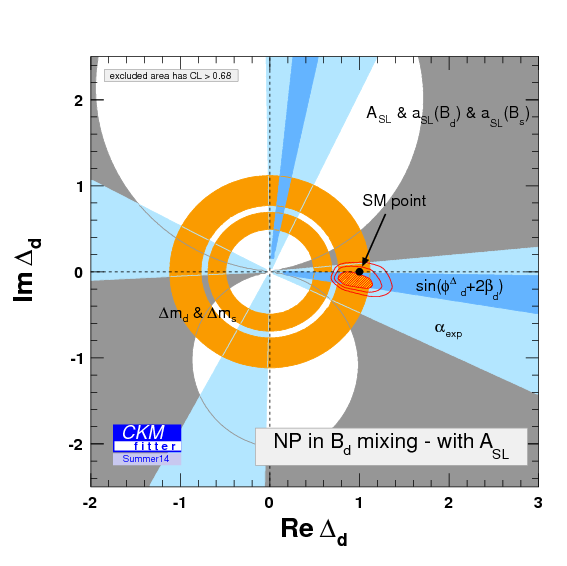}\includegraphics[width=0.45\linewidth]{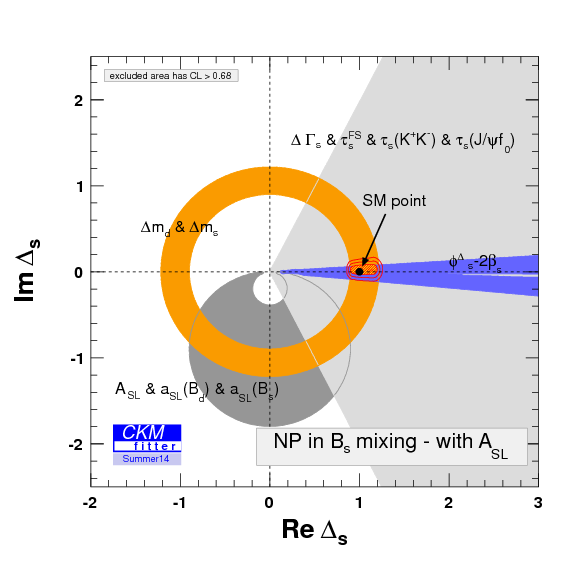}
\end{center}
\caption[]{Experimental determination of $\Delta_d$ from $B_d$ mixings (left panel) and $\Delta_s$ from $B_s$ mixings (right panel) by the CKMfitter collaboration \cite{CKMfitter2014}. The SM corresponds to $\Delta_{d,s}=1$.}
\label{fig:dec2}
\end{figure}
\begin{figure}[t!]
\begin{center}
\includegraphics[width=0.6\linewidth]{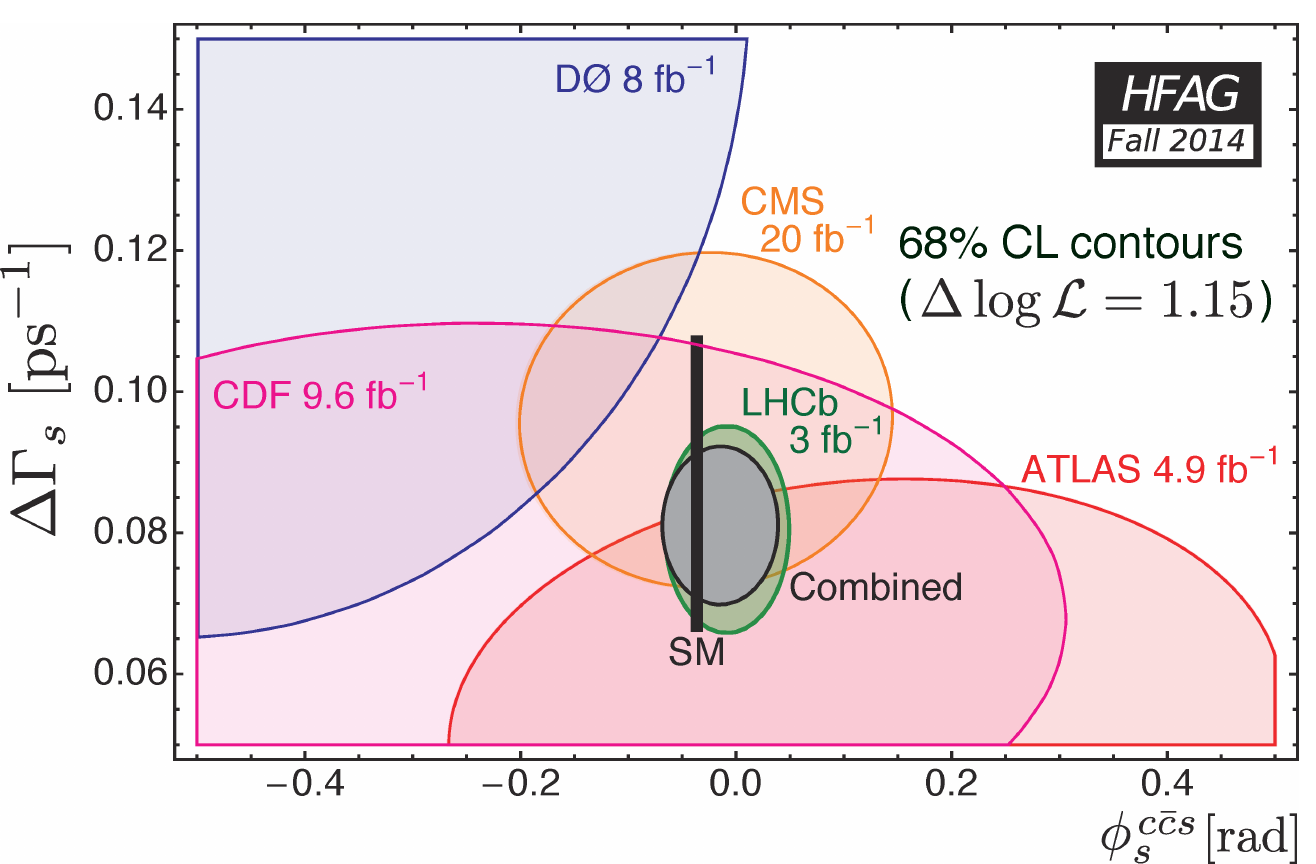}
\end{center}
\caption[]{Decay rate difference $\Delta\Gamma_s$ as a function of the new CP phase $\phi_s$ from analyses of the $b \to c {\bar c} s$ decays~\cite{Amhis:2014hma}.}
\label{fig:dec3}
\end{figure}
To summarise CP violation in the $b$ sector, Fig.~\ref{fig:dec2} shows the results of the CKMfitter collaboration for $B_d$ and $B_s$ mixings in Summer 2014. They correspond to a deviation by slightly more than 1$\sigma$ in $B_d$ mixings, and less than 1$\sigma$ in $B_s$ mixings. 
A combination of the results related to the decays of $b \to c {\bar c} s$ has also been performed by HFAG. Their results are presented in Fig.~\ref{fig:dec3}, which shows a good agreement with the SM, and that $\phi_s$ is compatible with the absence of new source of CP violation.

\subsection{CP violation in charm physics}

CP violation in charm physics is also very important, especially because it involves CKM matrix elements without CP violation.
Hence, CP violation in the charm sector is expected to be very small. The idea is similar to CP violation in the $b$ sector, where the $B_{(s)}$ mesons are replaced by $D_{(s)}$ mesons.
CP violation in charm physics can be probed in both $D_{(s)}$ mixings and $D_{(s)}$ decays. For the decays, the typical channels are $D^0 \to K^+ K^-$ and $\bar D^0 \to K^+ K^-$,
which probe $V_{cs}$ and $V_{us}$ and involve the $D^0 - \bar{D^0}$ oscillation (so time-dependent).
There are many other channels with 2, 3 or 4 light mesons (pions or kaons).
Another interesting observable that can be considered for $D_0$ decays is the difference of CP violation between $D^0\to K^+K^-$ and $D^0 \to \pi^+\pi^-$, defined as:
\begin{equation}
 \Delta A_{CP} \equiv A_{CP}(K^+ K^-) - A_{CP}(\pi^+ \pi^-)\;,
\end{equation}
which is expected to be small.

\begin{figure}[t!]
\begin{center}
\includegraphics[width=0.6\linewidth]{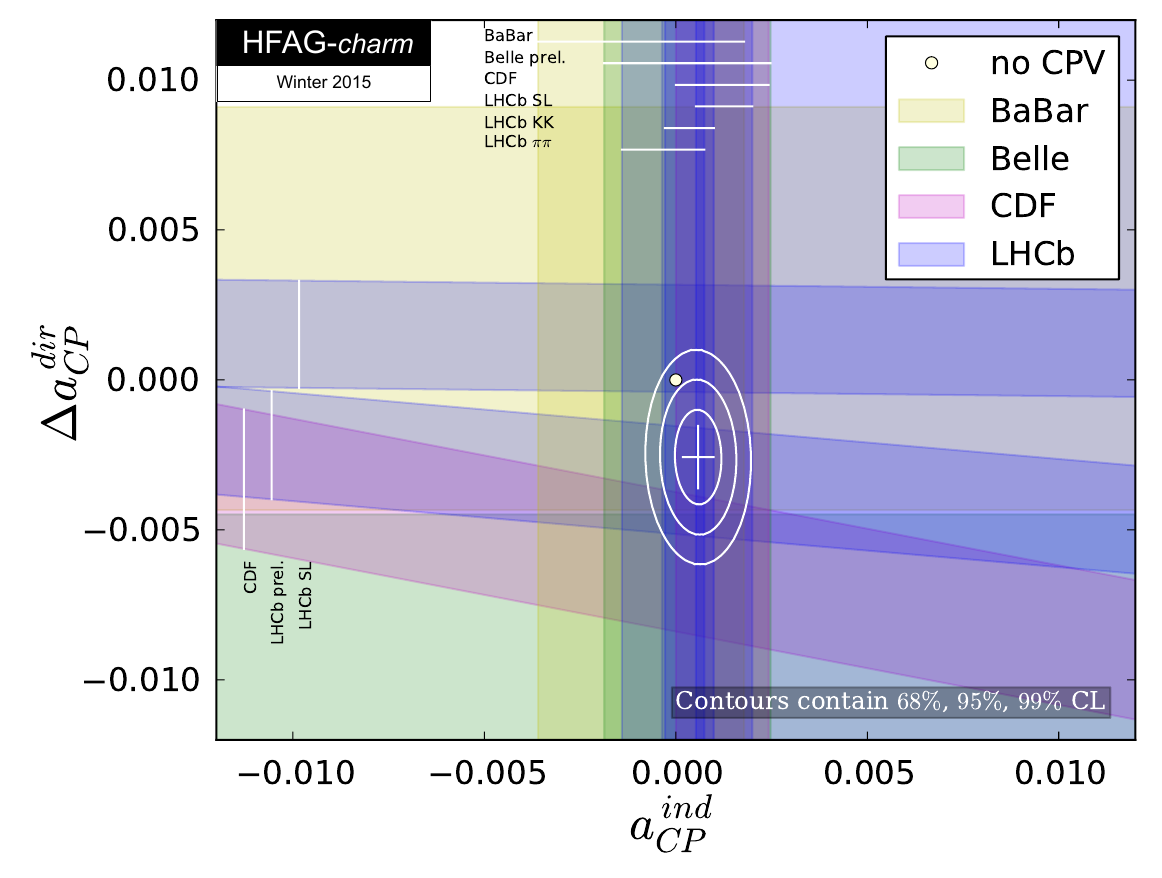}
\end{center}
\caption[]{$\Delta A_{CP}$ as a function of $A_{CP}$ in the charm sector~\cite{Amhis:2014hma}.}
\label{fig:charm2}
\end{figure}

Fig.~\ref{fig:charm2} presents $\Delta A_{CP}$ as a function of $A_{CP}$ using the combination of all the measurements from CDF, Babar, Belle and LHCb. The figure shows that $A_{CP}$ is compatible with 0, while $\Delta A_{CP}$ is slightly smaller than expected.

\section{Rare decays}

Another way to search indirectly for New Physics is through rare decays, which occur at loop level in the SM and are therefore very sensitive to NP effects. The theoretical framework for the calculation of rare decays is based on the effective field theory approach where the short distance (Wilson coefficients) and long distance (local operators) contributions are separated using Operator Product Expansion~\cite{Buras:1998raa}.

There have been several breakthroughs during the past few years, in particular with the first measurement of the $B_s \to \mu^+\mu^-$ branching ratio, and the measurement of clean angular observables in the $B\to K^* \mu^+\mu^-$ decay.

The branching ratio of $B_s \to \mu^+\mu^-$ can be calculated using
\begin{eqnarray}
\mathrm{BR}(B_s \to \mu^+ \mu^-) &=& \frac{G_F^2 \alpha^2}{64 \pi^3} f_{B_s}^2 \tau_{B_s} m_{B_s}^3 |V_{tb}V_{ts}^*|^2 \sqrt{1-\frac{4 m_\mu^2}{m_{B_s}^2}}\\
&& \times \left\{\left(1-\frac{4 m_\mu^2}{m_{B_s}^2}\right) \left|C_{S} -C'_{S} \right|^2 + \left | (C_{P} -C'_{P}) + 2 \, (C_{10} -C'_{10}) \frac{m_\mu}{m_{B_s}} \right |^2\right\}\nonumber\;,
\end{eqnarray}
where $C_{10}$ embeds the SM contribution, and $C_S$ and $C_P$ are scalar and pseudoscalar coefficients which can receive large contributions from NP. The $C'_i$ denote the chirality flipped Wilson coefficients. $f_{B_s}$ is the $B_s$ decay constant which constitutes the largest source of uncertainty. The SM prediction for this branching ratio is $(3.54 \pm 0.27) \times 10^{-9}$, based on~\cite{Mahmoudi:2008tp,Hermann:2013kca,Bobeth:2013tba}, 
which is in agreement with the combined CMS and LHCb measured value of $(2.8^{+0.7}_{-0.6})\times 10^{-9}$ presented in~\cite{CMS:2014xfa}. The compatibility between the SM values and the experimental measurement sets strong constraints on New Physics models, in particular on supersymmetry~\cite{Arbey:2012ax}, where the scalar and pseudoscalar contributions are enhanced approximately as $\tan^6\beta/M_A^4$.

\begin{figure}[t!]
\begin{center}
\includegraphics[width=0.5\linewidth]{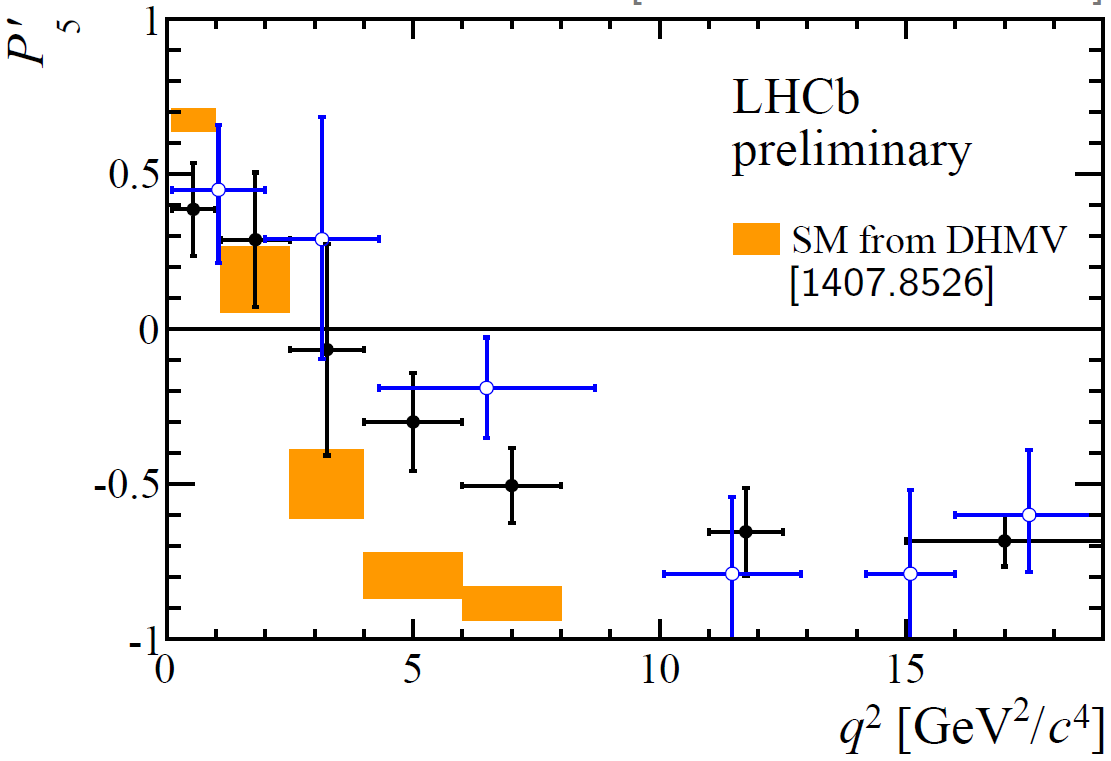}
\end{center}
\caption[]{LHCb experimental measurements of the $P'_5$ observable~\cite{LHCb:2015dla} with 3 fb$^{-1}$ of data (in black), compared to the 1 fb$^{-1}$ results (in blue) and the theoretical predictions~\cite{Descotes-Genon:2014uoa}, as a function of $q^2$.}
\label{fig:p5}
\end{figure}

The $B\to K^* \mu^+\mu^-$ decay also provides a multitude of observables sensitive to different helicity structures in the decay amplitude. Unfortunately, the theoretical predictions for the usual observables inherit large uncertainties from the hadronic form factors. This has led to the construction of a number of optimised observables as appropriate ratios of angular coefficients where the form factor uncertainties cancel at leading order, while having high sensitivity to NP effects~\cite{DescotesGenon:2012zf}. LHCb has found a 2.9$\sigma$ discrepancy with the SM predictions in two of the $q^2$ bins for one of these clean angular observables~\cite{LHCb:2015dla}, namely in the bins $q^2 \in [4.0, 6.0]$ and $[6.0, 8.0]$ GeV$^2$ of the observable $P'_5$, as can be seen in Fig.~\ref{fig:p5}. 

\begin{figure}[h!]
\begin{center}
\includegraphics[width=0.45\linewidth]{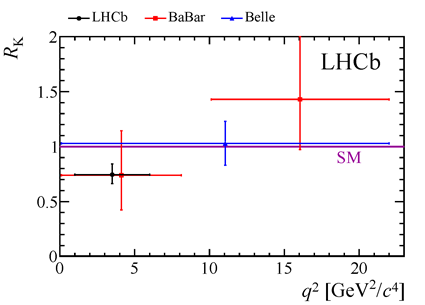}
\end{center}
\caption[]{LHCb, Babar and Belle experimental measurements of the $R_K$ ratio, as a function of $q^2$~\cite{Aaij:2014ora}.}
\label{fig:RK}
\end{figure}

Another recent discrepancy measured by LHCb is the ratio $R_K$ of the branching ratio of $B\to K \mu^+\mu^-$ over the one of $B\to K e^+e^-$ for $q^2 \in [1,6]$ GeV$^2$, which is a probe of lepton universality. This ratio is expected to be close to 1 in the SM~\cite{Hiller:2003js}, with uncertainties lower than 1\%. The LHCb result is $0.745^{+0.090}_{-0.074} ({\rm stat.}) \pm 0.036 ({\rm syst.})$~\cite{Aaij:2014ora}, showing a deficit of 
about 25\%. This result is compared to the ones from Belle and Babar in Fig.~\ref{fig:RK}.

\begin{figure}[t!]
\begin{center}
\includegraphics[width=0.45\linewidth]{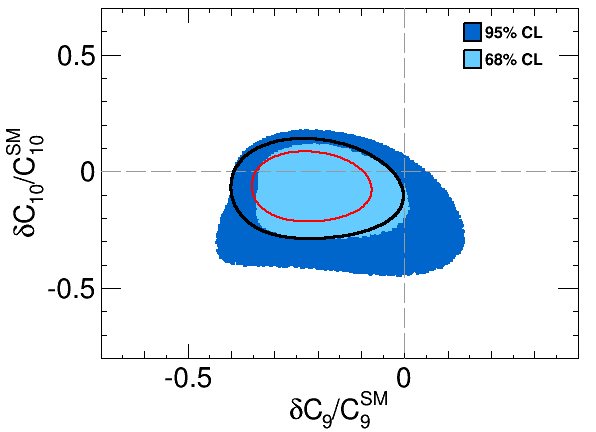}\hspace*{0.5cm}\includegraphics[width=0.45\linewidth]{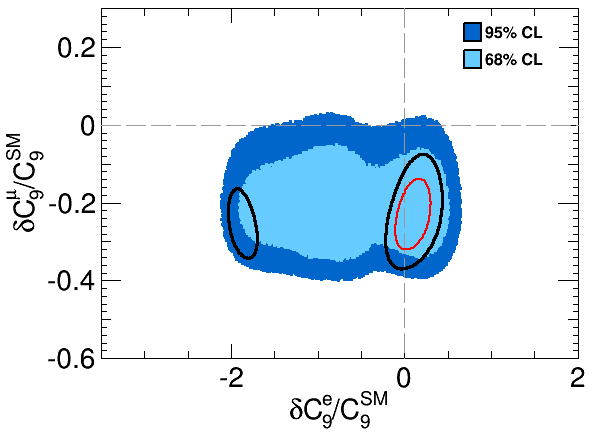}
\end{center}
\caption[]{Global fits to the $b\to s\ell\ell$ observables~\cite{Hurth:2014vma}. $\delta C_i$ corresponds to the NP contribution to the Wilson coefficient $C_i$. The light (dark) blue zone shows the 68\% (95\%) C.L. region. In the left panel, the result of a global fit to $C_9,C_{10},C_9',C_{10}'$ is compared to the result obtained if only $C_9,C_{10}$ are varied (red and black contours). In the right panel, the result of a global fit to $C_9^\mu,C_9^e,C_9^{\prime\mu},C_9^{\prime e}$ is compared to the result obtained if only $C_9^\mu,C_9^e$ are varied (red and black contours).}
\label{fig:c9}
\end{figure}

Discrepancies in both $B\to K^* \mu^+\mu^-$ and $B\to K \ell^+\ell^-$ may be related to New Physics but a global fit of all the $b\to s\ell\ell$ observables shows that the discrepancy is smaller than 2$\sigma$ if four Wilson coefficients or more are allowed to vary \cite{Hurth:2014vma}. This is examplified in Fig.~\ref{fig:c9}, where in the left panel a global fit to the Wilson coefficients $C_9,C_{10},C_9',C_{10}'$ is compared to the results obtained if only $C_9,C_{10}$ are varied. Similarly, in the right panel, a global fit to the Wilson coefficients $C_9^\mu,C_9^e,C_9^{\prime\mu},C_9^{\prime e}$ is shown and compared to the fit to $C_9^\mu,C_9^e$. 

While the current discrepancies are for the moment not significant enough to claim for any discovery, it is interesting to notice that a modification of the Wilson coefficient $C_9^\mu$
seems to provide a coherent answer favoured by the current results (see for example Refs.~\cite{Descotes-Genon:2013wba,Hiller:2014yaa}).

\section{Conclusions}

Heavy flavour physics plays a major role in our understanding of the fundamental questions in particle physics. Several decades of $B$ factory measurements and recently also the LHC experiments have provided impressive tests of the SM paradigms and parameters. More recent challenges of flavour physics are focused on finding indirect paths to New Physics, mainly through CP violating and rare decay observables. At the moment, the current experimental data do not point to new source of CP violation. On the other hand, there exist a few deviations with the SM predictions in the semi-leptonic rare decays, although not significant enough to be conclusive yet. The impressive progress in the theoretical calculations and lattice results in the recent years have been crucial in this context. The next runs of the LHC and a future high luminosity $B$ factory are likely to provide more insight to settle the current discrepancies or point to New Physics phenomena.

\section*{Acknowledgments}

The author is grateful to the organisers of Moriond QCD 2015 for the fruitful conference and for their invitation, and acknowledges partial support from Institut Universitaire de France.

\end{document}